\documentstyle[12pt,preprint]{aastex}

\received{June 14, 2002}
\accepted{August 6, 2002}

\def\as {{$^{\prime\prime}$}}
\def\am {{$^\prime$}}
\def\deg{{$^\circ$}}

\begin{document}

\title{Abundances of Molecular Species in Barnard 68}

\author{James Di Francesco\altaffilmark{1}, Michiel R. Hogerheijde\altaffilmark{2}, William J.
Welch}
\affil{Radio Astronomy Laboratory, 601 Campbell Hall, \\ 
University of California, Berkeley, Berkeley, CA, \\
94705-3411 U.S.A.}
\author{Edwin A. Bergin}
\affil{Harvard-Smithsonian Center for Astrophysics \\
60 Garden Street, MS 42, Cambridge, MA 02138, U.S.A.}
\altaffiltext{1} {currently at Herzberg Institute of Astrophysics, National 
Research Council of Canada, 5071 West Saanich Road, Victoria, BC V9E 2E7 
Canada}
\altaffiltext{2} {currently at Steward Observatory, University of Arizona,
933 North Cherry Avenue, Tucson, AZ 85721-0065, U.S.A.}

\begin{abstract}
Abundances for 5 molecules (C$^{18}$O, CS, NH$_{3}$, H$_{2}$CO, and 
C$_{3}$H$_{2}$) and 1 molecular ion (N$_{2}$H$^{+}$) and upper limits 
for the abundances of 1 molecule ($^{13}$CO) and 1 molecular ion 
(HCO$^{+}$) are derived for gas within the Bok globule Barnard 68 
(B68).  The abundances were determined using our own BIMA millimeter 
interferometer data and single-dish data gathered from the literature, 
in conjunction with a Monte Carlo radiative transfer model.  Since B68 
is the only starless core to have its density structure strongly 
constrained via extinction mapping, a major uncertainty has been 
removed from these determinations.  All abundances for B68 are lower 
than those derived for translucent and cold dense clouds, but perhaps 
only significantly for N$_{2}$H$^{+}$, NH$_{3}$, and C$_{3}$H$_{2}$.
Depletion of CS toward the extinction peak of B68 is hinted at by 
the large offset between the extinction peak and the position of 
maximum CS line brightness.  Abundances derived here for C$^{18}$O and 
N$_{2}$H$^{+}$ are consistent with other, recently determined values 
at positions observed in common.
\end{abstract}


\clearpage
\section{Introduction}

Stars form out of cores of dense molecular gas.  Such cores can be embedded 
deeply within molecular clouds, but numerous examples of isolated cores, e.g., 
Bok globules (Bok \& Reilly 1947; Clemens \& Barvainis 1988), have been also
identified as dark patches of high visual extinction against rich stellar
backgrounds.  As isolated cores, Bok globules can be relatively free from the 
influences of nearby star-forming events, and so can provide very simple and 
pristine examples of the physical and chemical structures of cores prior 
to the formation of low-mass stars.  In this paper, we examine the abundances 
of several species of molecules or molecular ions in the globule Barnard 
68 (B68; Barnard 1919). 

B68 (also known as CB82 or LDN57) is an Ophiuchus dark cloud, located at 
$\sim$125~pc (de Geus, de Zeeuw, \& Lub 1989)\footnote{A distance range of
60-205 pc for the Ophiuchus dark clouds was formally estimated by de Geus 
et al. with 125 pc $\pm$ 25 pc defined as their center.}, within the Loop 
I superbubble.  Figure 1 shows an $R$-band image of B68 obtained from the 
CDS/Aladin sky atlas, revealing its compact, roundish morphology.  B68 has 
no evidence of star formation within its interior, e.g., from detections 
of outflow wings in lines of $^{12}$CO (Avery et al. 1987) or from an IRAS 
point source (Parker 1988)\footnote{The 24\as\ $\times$ 5\as\ (P.A. of 
94\deg) error ellipse of the IRAS point source 17194-2351, associated with 
B68 by Clemens \& Barvainis, is centered $\sim$3\am\ beyond the opaque edge 
of B68 seen in Figure 1.}.  Furthermore, continuum emission at $\lambda$ = 
1.3 mm toward B68 has not been detected (Reipurth, Nyman, \& Chini 1996;
Launhardt \& Henning 1997).  

B68 is unique in that it is the only starless Bok globule to have its 
density structure well defined.  Using the extinction of the background 
Galactic bulge K-giant population, Alves, Lada, \& Lada (2001; ALL01) 
found an azimuthally-averaged radial column density profile for B68 that 
was well matched with that of an isothermal, self-gravitating Bonnor-Ebert 
sphere, specifically one with a near-critical center-to-edge density 
contrast of 16.5 (Bonnor 1956; Ebert 1955).  The robust definition of 
density within B68 removes a major uncertainty towards determining its 
molecular abundances.  Recently, Bergin et al.\/ (2002) used single-dish 
maps of B68 in conjunction with models specifically incorporating the 
ALL01 Bonnor-Ebert density structure to ascertain radial variations in 
C$^{18}$O and N$_{2}$H$^{+}$ abundance, with the lowest values occurring 
at the highest extinctions within B68.  Hotzel et al.\/ (2002) also found 
evidence for depletion of CO isotopes with extinction from their own 
single-dish maps and a re-analysis of the ALL01 data.  Each group 
speculated these molecules sublimate from the gas phase onto dust grains 
at the high densities and low temperatures in the B68 interior.  A similar 
interpretation was made by Tafalla et al.\/ (2002) to explain the radial 
depletion of CS and C$^{18}$O in the interiors of 5 other starless cores, 
although N$_{2}$H$^{+}$ was not depleted and NH$_{3}$ was enhanced at 
these locations.

We explore further the abundances of several molecules or molecular ions 
within B68, incorporating the ALL01 Bonnor-Ebert density structure into our 
own models.  New observations of B68 made with the BIMA millimeter array are 
used to obtain upper limits on abundances of $^{13}$CO and HCO$^{+}$ near 
the extinction peak found by ALL01, i.e., at R.A./decl. (2000) =  
17$^{h}$22$^{m}$38.6$^{s}$, -23$^{\circ}$49$^{\prime}$46.0$^{\prime\prime}$.  
Upper limits to C$^{18}$O and N$_{2}$H$^{+}$ abundances are also derived at 
this position, consistent with values derived by Bergin et al.  In addition, 
single-dish detections of lines reported in the literature are used to derive 
abundance values for C$^{18}$O, CS, N$_{2}$H$^{+}$, NH$_{3}$, H$_{2}$CO, and
C$_{3}$H$_{2}$ toward positions in B68 offset from the extinction peak.  We 
describe the BIMA observations and single-dish data in \S 2, and our method 
of determining abundances with a Monte Carlo radiative transfer code in \S 3.  
The derived results are discussed in \S 4.  A concluding summary is found in 
\S 5.

\section{Observations and Results}

\subsection{BIMA Observations}

Figure 1 shows the position of the one pointing made toward B68 with the 
BIMA millimeter interferometer at Hat Creek, CA, i.e., at R.A./decl. (2000) 
= 17$^{h}$22$^{m}$38.6$^{s}$, -23$^{\circ}$49$^{\prime}$42.0$^{\prime\prime}$, 
approximately 4\as\ north of the extinction peak.  Both $^{13}$CO 1--0 and 
C$^{18}$O 1--0 were observed simultaneously during tracks on 2001 May 09 and 
May 20, and N$_{2}$H$^{+}$ 1--0, HCO$^{+}$ 1--0, and HOC$^{+}$ 1--0 were 
observed simultaneously on 2001 April 24 and May 14.  Figure 1 also shows 
the minimum and maximum FWHM sizes of the BIMA primary beams at the 
frequencies of $^{13}$CO 1--0 and HCO$^{+}$ 1--0.  Each line was observed 
in its own correlator window of 6.25 MHz width, with each window set to have 
512 channels of 12.2 kHz width (i.e., $\sim$0.03 km s$^{-1}$ at 110 GHz.)  
The correlators were tuned so the central channel of each window corresponded 
to a $V_{LSR}$ of $+$3.4 km s$^{-1}$, the velocity centroid of C$^{18}$O 2--1 
found by Wang et al. (1995).

Each track was only 4-5 hours in duration given the low maximum elevation of 
B68 from Hat Creek (i.e., $\sim$25\deg).  Visibility phases were calibrated 
by observing 1733-130 for 5 minutes approximately every 30 minutes over 800 
MHz bandwidth in the lower or upper sideband.  Visibility amplitudes were 
calibrated by similarly observing Uranus for 8 minutes at the end of each 
track.  All data were reduced using standard routines within the MIRIAD 
software package (Sault, Teuben, \& Wright 1995).  All channels were cleaned 
to 2 $\sigma$ rms intensity levels.

Line emission was not detected at any of the frequencies observed, at any
positions toward B68 within the $\sim$2\am\ FWHMs of the BIMA primary beams.  
Binning together several channels or tapering the data with variously-sized 
Gaussians did not change this result.  Table 1 summarizes these data, listing 
the lines, the source observatory, the synthesized beam FWHMs, and the angular 
offset of the BIMA pointing from the extinction peak.  The 1 $\sigma$ rms 
sensitivities attained per channel ranged from 1.0~K for HOC$^{+}$ 1--0 to 
1.5~K for $^{13}$CO 1--0, and were measured by sampling all channels within 
a 40\as\ $\times$ 40\as\ box centered at the pointing center.  Upper limits 
(at 3 $\sigma$) to the integrated intensities of these lines ranged from 
0.21~Jy~beam$^{-1}$~km~s$^{-1}$ to 0.34~Jy~beam$^{-1}$~km~s$^{-1}$ (or 
0.25~K~km~s$^{-1}$ to 0.44~K~km~s$^{-1}$) assuming a line width equal to 
0.4~km~s$^{-1}$, the typical FWHM of optically-thin lines in B68 found by 
Wang et al.

\subsection{Single-dish Observations}

We surveyed the literature to find previous detections of molecular line
emission toward B68 from which abundances could be measured.  Fortunately, 
B68 was included in numerous surveys of dark globules for molecular line 
emission.  Early detections of lines include those of $^{12}$CO 1--0 and 
$^{13}$CO 1--0 by Martin \& Barrett (1978) and Leung, Kutner, \& Mead (1982), 
$^{12}$CO 2--1 by Avery et al.\/ (1987) and Clemens, Yun, \& Heyer (1991), 
and $^{12}$CO 3--2 by Avery et al.  Furthermore, NH$_{3}$ (1,1) and (2,2) 
were detected by Martin \& Barrett and Bourke et al.\/ (1995).  Although 
these data certainly associate line emission with B68, we exclude them 
from analysis because of their relatively low angular resolution, i.e., 
$>$60\as\ FWHM.  

More recent detections of line emission from various species toward B68 have
been made with resolutions $<$60\as\ FWHM, including those of C$^{18}$O 2--1 
and H$_{2}$CO 3$_{12}$--2$_{11}$ by Wang et al.\/, CS 2--1 by Launhardt et 
al.\/ (1998), N$_{2}$H$^{+}$ 1--0 and C$_{3}$H$_{2}$ 2$_{02}$--1$_{01}$ by 
Benson, Caselli, \& Myers (1998), and NH$_{3}$ (1,1) by Lemme et al.\/ (1996).
Other recent detections include those of C$^{18}$O 1--0 and N$_{2}$H$^{+}$ 
1--0 by Bergin et al.\/, and $^{13}$CO 1--0, C$^{18}$O 1--0 and 2--1 by 
Hotzel et al., but we exclude those data from our analysis because abundances 
have been specifically determined from them by the respective authors.  An 
interesting non-detection in the literature is that of SO 1$_{0}$--0$_{1}$ 
by Codella \& Muders (1997), but their map was centered on the nearby IRAS 
point source and did not overlap the region of extinction modeled by ALL01.  
The methods of observation and reduction relevant to these data are described 
in these references, and are not reproduced here.  

In the higher-resolution single-dish studies, the positions where line 
characteristics were specifically reported vary widely.  Figure 1 shows 
the positions and resolutions of observations from these studies against 
the $R$-band image of B68.  Table 1 also lists the lines observed, the 
source observatories, the resolutions attained, and the angular offsets 
of the single-dish data from the extinction peak of the reported line 
characteristics.  The positions listed in Table 1 for C$^{18}$O 2--1, 
CS 2--1, and N$_{2}$H$^{+}$ 1--0 are those of peak line intensity from 
their respective maps, where line characteristics were specifically 
reported.  (H$_{2}$CO 3$_{12}$--2$_{11}$ and C$_{3}$H$_{2}$ 2$_{02}$--1$_{01}$ 
were observed at only these positions by the respective authors.)  
Interestingly, these positions are not coincident with the extinction peak.  
The C$^{18}$O and CS positions are approximately coincident with an arc 
of maximal C$^{18}$O 1--0 integrated intensity centered at the 
extinction peak noted by Bergin et al.\/ (see their Figure 1a), and the 
N$_{2}$H$^{+}$ position is approximately coincident with the ``arc" of 
maximal N$_{2}$H$^{+}$ 1--0 integrated intensity about the extinction peak 
also noted by Bergin~et~al. (see their Figure 1b.)\footnote{An N$_{2}$H$^{+}$ 
1--0 map toward B68 recently made by Caselli et al.\/ (2002), confirms this 
position is that of peak intensity, although their data are not Nyquist 
sampled and are of lower resolution than those of Benson et al.\/ and 
Bergin et al.}

\section{Monte-Carlo Radiative Transfer Models} 

Molecular line data can place strong limits on molecular abundances within 
B68 because its density structure has been so well constrained by ALL01.  
For example, abundances can be estimated by calculating with a Monte 
Carlo code the radiative transfer, molecular excitation, and line emission 
through a model of a dense core with a specific density profile, and varying 
abundances within the model until the output matches the observed data.  
For this purpose, we used the one-dimensional Monte Carlo code of Hogerheijde 
\& van der Tak (2000) which can solve molecular excitation coupled with 
line and continuum radiative transfer in core models.  This code is 
especially useful for application to B68.  First, model fluxes can be found 
at positions arbitrarily offset from the central line-of-sight, to simulate 
observations of B68 at positions offset from the extinction peak.  Second, 
the model fluxes can be sampled with the spatial frequency coverage of 
actual interferometer data to estimate how much emission may have been 
resolved out in those cases.  Recently, van Zadelhoff et al. (2002) 
compared 8 molecular excitation/radiative transfer codes including that 
of Hogerheijde \& van der Tak, and found that all codes give results to 
within 10\% of one another, even at high opacity.

For models of B68, we assumed the thermal and density structure given 
by ALL01, i.e, an isothermal sphere with an outer radius of 12 500 AU 
and a ``Bonnor-Ebert parameter" $\xi_{\rm max}$ of 6.9, corresponding
to a center-to-edge density ratio of 16.5.  Following ALL01, the 
temperature assumed in our models was 16 K, as derived by Bourke et 
al.\/ from NH$_{3}$ observations of B68.  (An alternative temperature 
profile is considered in \S 4.1 below.)  The $\xi_{\rm max}$ parameter 
identifies the particular solution from the family that solves the 
second-order differential equation characterizing a Bonnor-Ebert sphere.  
We determined the density in 30 linearly-spaced concentric shells by 
solving that differential equation with a fourth-order Runge-Kutta 
method (see Press et al.\/ 1992.)  The central density in the ALL01
model for B68 is $\sim$2.5 $\times$ 10$^{5}$ cm$^{-3}$.

Molecular abundances were assumed to be constant with radius, although 
abundance gradients within B68 are quite possible for the molecular species 
considered here.  For example, Bergin et al.\/ found edge-to-center 
contrasts in the abundances of C$^{18}$O and N$_{2}$H$^{+}$ of 100 and 2 
respectively using their own maps of B68.  With only single pointings 
for the single-dish data and non-detections for the interferometer data, 
we can derive abundance values or upper limits only at the positions 
listed in Table 1, averaged over the respective lines-of-sight and beam 
widths, and cannot probe for abundance gradients.  Gradients should not 
produce dramatic differences in the abundances we derive, however, given 
the small sizes of the beam widths relative to the spatial extent of B68.  
For example, we note in \S 4.2 little difference between our abundances 
of C$^{18}$O and N$_{2}$H$^{+}$ and those derived by Bergin et al.\/ at
positions observed in common, despite our different abundance profiles.

The velocity field of the gas was assumed to have a turbulent line width 
of 0.4 km s$^{-1}$ (see Wang et al.\/; since the lines are optically thin 
at the abundances found, the integrated intensity is independent of the 
adopted turbulent line width.)  We assume that each shell is stationary, 
although Bergin et al.\/ suggest velocity variations within B68 may be 
significant.  For reasons similar to those cited above for abundance 
gradients, we did not consider radial variations of line width.

In each shell, level populations were determined assuming standard collision 
rates for each molecule.  The expected sky brightness distributions were 
then calculated using a $256 \times 256 \times 100$ cube with $1''\times 1'' 
\times 0.04$ km s$^{-1}$ elements.  To compare models with interferometer 
data, ``visibility" datasets were produced from the cubes using the MIRIAD 
task {\it uvmodel}\/ and the antenna baselines from the original data.  
These latter datasets were inverted, cleaned and restored in the same manner 
as the actual data, and velocity-integrated line intensities calculated.  
By properly accounting for spatial filtering in this manner, we estimated 
BIMA would have recovered only $\sim$20\% of the flux emitted from an ALL01 
Bonnor-Ebert sphere in B68.  To compare models with single-dish data, the 
original cubes were convolved with Gaussians with widths appropriately 
representing the resolution of the respective observations.  The abundances 
were varied until the observational data were well matched.  

\section{Abundances}

Table 2 lists the fractional abundances of the sampled molecular species
at various positions toward B68, constrained with the data described in 
\S 2, and estimated by using the method described in \S 3 (assuming models 
with the Bonnor-Ebert density profile of ALL01.)  The BIMA data yield the 
first upper limits to the abundances of HCO$^{+}$ toward the extinction 
peak of B68, as well as upper limits to the abundances of $^{13}$CO, 
C$^{18}$O and N$_{2}$H$^{+}$ at the same location.  Since collision rates 
are not available for HOC$^{+}$, we do not attempt to place limits on its 
abundance.  In addition, single-dish data of B68 from the literature yield 
abundance values of H$_{2}$CO, CS, NH$_{3}$, and C$_{3}$H$_{2}$ toward B68 
for the first time, but at various positions offset from the extinction 
peak.  Finally, other single-dish data of B68 from the literature yield 
new abundance values of C$^{18}$O and N$_{2}$H$^{+}$ at some of these 
latter locations.

\subsection{Comparison with Other Clouds}

How do the abundances derived for B68 here compare to those derived 
for other clouds?  ``Standard" cloud molecular abundances with which 
to base comparisons are difficult to define given wide variations in 
chemical environment or evolutionary epoch in the ISM.  To provide some 
comparison, Table 2 also lists the fractional abundance estimates of the 
same species in clouds similar in character to B68, obtained directly 
from the literature but made without the detailed foreknowledge of cloud 
density structure now available for B68.  The first set, for ``translucent 
clouds," are those listed by Turner (2000) for small, round Clemens \& 
Barvainis clouds with edge-to-center visual extinctions of 2.0.  These
clouds were modeled with constant abundance, and the listed values are 
averages between the results of a hydrostatic equilibrium polytropic 
model and an $n$($r$) $\propto$ $r^{0}$ model\footnote{According 
to Turner (2002, private communication), these two models produce results 
that differ at most by a factor of 1.35 for species that are highly 
dependent on density.}.  The second set, denoted for ``cold dense clouds," 
are those compiled by Ohishi, Irvine, \& Kaifu (1992) for a single position 
in TMC-1 or L134N, and were made using column density estimates from many 
authors, assuming N(H$_{2}$) = 10$^{22}$~cm$^{-2}$ (see Turner, or Pratap 
et al.\/ 1997, or Dickens et al.\/ 2000 for alternative estimates.)  The
abundances from Turner and Ohishi et al.\/ in Table 2 should not be regarded 
as universal.  For example, Turner notes abundances can vary by an order 
of magnitude both between different translucent clouds and within larger, 
dense clouds (presumably determined using similar assumptions.) 

Table 2 shows that every value of molecular abundance we derive for B68 is 
less than the lowest value derived for other clouds.  From the BIMA data, 
we find the upper limits to the $^{13}$CO, HCO$^{+}$, and N$_{2}$H$^{+}$ 
abundances in B68 are lower than values derived for other clouds by factors 
of 7, 14, and 4 respectively.  (The upper limit to the C$^{18}$O abundance 
is similar to the lowest value found for other clouds.)  From the various
single-dish data, we find the C$^{18}$O, CS, and H$_{2}$CO abundances differ 
least, by roughly an order of magnitude or less, with the B68 abundances 
lower than the lowest values derived for other clouds by factors of 
only 5, 3, and 16 respectively at the locations observed.  However, the 
N$_{2}$H$^{+}$, NH$_{3}$, and C$_{3}$H$_{2}$ abundances differ the most, 
by over an order of magnitude, with those of B68 lower by factors of 25, 
29, and 170 respectively at the locations observed.  

The assumption in our models of an isothermal Bonnor-Ebert sphere, at 
the same 16~K temperature derived by Bourke et al.\/ and used by ALL01, 
has likely minimized the resulting abundance values we obtain.  Other 
temperature profiles, such as those suggested for Bonnor-Ebert spheres 
by Zucconi, Walmsley, and Galli (2001; see also Evans et al.\/ 2001)
which have cooler temperatures at small radii, may be more appropriate.  
For example, Bergin et al.\/ assumed for their models of B68 the Zucconi 
et al.\/ radial temperature profile with a global reduction of 2~K, in 
addition to the ALL01 density profile.  With this same temperature 
profile in our models, the values in Table 2 increase by factors of 
only 2--3.  Therefore, using this radial temperature profile would 
bring the abundances of C$^{18}$O, CS, and H$_{2}$CO at the positions 
observed in B68 even more in line with the abundances found for other 
cloud types.  However, the abundances of N$_{2}$H$^{+}$, NH$_{3}$, and 
C$_{3}$H$_{2}$ at other positions would still remain lower than those of 
other clouds by an order of magnitude or more.

Only N$_{2}$H$^{+}$, NH$_{3}$, and C$_{3}$H$_{2}$ are arguably depleted 
in B68, given the values shown in Table 2 and assuming a typical order of 
magnitude abundance variation between and within similar clouds.  However, 
other species considered here still may be depleted in B68.  Note that 
the degree to which the abundance of a particular species in B68 is low 
compared to those of other clouds appears related to the position across 
B68 that the respective data were obtained.  The C$^{18}$O, CS, and 
N$_{2}$H$^{+}$ data from the literature were those of maximum line 
brightness from maps of B68, and H$_{2}$CO and C$_{3}$H$_{2}$ data were 
obtained only subsequently by the respective authors at those positions.
(NH$_{3}$ was observed only at the Clemens \& Barvainis position of 
B68.)  Figure 1 and Table 2 together reveal that the less-discrepant 
abundances are found at positions relatively far from the extinction 
peak but the more-discrepant abundances are found at positions closer 
to the extinction peak (except notably NH$_{3}$.)  This pattern suggests
C$_{3}$H$_{2}$ may be also depleted by some process related to extinction, 
e.g., the sublimation of gas-phase molecules onto grains, as suggested by 
Bergin et al.\/ for C$^{18}$O and N$_{2}$H$^{+}$ and Hotzel et al.\/ for 
$^{13}$CO and C$^{18}$O.  This same idea may also explain how CS and 
H$_{2}$CO appear relatively undepleted in the outer, less-extincted radii 
of B68.  Moreover, the lack of bright emission in CS near the extinction 
peak hint it may be also depleted at high extinction in B68.  These 
speculations can be confirmed only after analyzing fully-sampled, 
high-resolution line maps of B68.

\subsection{Comparison with Previous Abundance Determinations}

How do the abundances we derive compare with those found previously for
B68?  Bergin et al.\/ found the C$^{18}$O abundance in B68 rises from 
very low values at $A_{V}$~$<<$~1 to a peak of 1~$\times$~10$^{-7}$ at 
$A_{V}$~=~2, and decreases to 1~$\times$~10$^{-9}$ at $A_{V}$~$>$~20,
a contrast of $\sim$100.  In addition, Bergin et al.\/ found the 
N$_{2}$H$^{+}$ abundance rises from very low values at $A_{V}$~$<<$~1 
to a peak of 6~$\times$~10$^{-11}$ at $A_{V}$~=~3, and decreases to 
3~$\times$~10$^{-11}$ at $A_{V}$ $>$~20, a contrast of $\sim$2.  Any 
evidence of N$_{2}$H$^{+}$ depletion is remarkable, given its 
oft-described utility as a non-depleting probe of dense core interiors 
(e.g., see Tafalla et al.\/ 2002.)  

Our interferometer data of C$^{18}$O and N$_{2}$H$^{+}$ do not provide much
additional support for the abundance model for B68 of Bergin et al.  Our 
derived upper limits for C$^{18}$O and N$_{2}$H$^{+}$ from the BIMA data 
are not particularly low, but they remain consistent with the still-lower 
values found by Bergin et al.\/ throughout the core.  Also, the non-detection 
of compact line emission with BIMA suggests the low abundances found by 
Bergin et al.\/ are not due to the single-dish beam dilution of small clumps 
of relatively abundant material.  A spatially-smoother model of gas density 
is indeed more appropriate for B68, such as the Bonnor-Ebert sphere suggested 
by the dust extinction maps.

The single-dish C$^{18}$O and N$_{2}$H$^{+}$ data from the literature only 
pertain to one line-of-sight per transition toward B68, but provide strong 
support for the abundances derived by Bergin et al., if the same modified 
Zucconi et al.\/ temperature profile is assumed.  With this assumption, 
the C$^{18}$O literature data, from a position 59\farcs4 offset from the 
extinction peak, yield an apparent abundance of 6--9 $\times$ 10$^{-8}$, 
quite consistent with the 3--9 $\times$ 10$^{-8}$ abundance range expected 
along this line-of-sight from the model of Bergin et al.  Also, the 
N$_{2}$H$^{+}$ literature data, from a position 16\farcs3 offset from the 
extinction peak, yield an apparent value of 4--6 $\times$ 10$^{-11}$, 
slightly larger than but still consistent with the 3--4 $\times$ 10$^{-11}$ 
abundance range expected along this line-of-sight from the model of Bergin 
et al.  Furthermore, the observed integrated intensities of these literature 
data at each respective location and our BIMA observations can be reproduced 
to within only a few percent with our own Monte Carlo models, assuming the 
ALL01 density profile, the modified Zucconi et al.\/ temperature profile 
and the Bergin et al.\/ radial abundance profiles.  (Note that here we 
relaxed the earlier assumption of constant abundance with radius.)

\section{Summary and Conclusions}

Using our own interferometer line data and single-dish line data culled 
from the literature of the Bok globule B68 and a Monte-Carlo radiative 
transfer code, we derive abundances for C$^{18}$O, CS, N$_{2}$H$^{+}$, 
NH$_{3}$, H$_{2}$CO, and C$_{3}$H$_{2}$ and upper limits for abundances 
of $^{13}$CO and HCO$^{+}$.  Foreknowledge of the Bonnor-Ebert density 
configuration of B68, as found from extinction mapping by ALL01, has 
removed a major uncertainty that affected previous estimations of 
abundances in other clouds.  We find molecular abundances in B68 are lower 
than those estimated previously for clouds similar to B68.  N$_{2}$H$^{+}$, 
NH$_{3}$, and C$_{3}$H$_{2}$ have abundances lower than the lowest values 
determined for other clouds by 1.4, 1.5, and 2.2 orders of magnitude 
respectively, but abundances of the other species are only lower by one 
order of magnitude or less.  Depletion of CS is suggested by the large 
offset between the position of maximum CS line brightness and the extinction 
peak.  Furthermore, abundances of C$^{18}$O and N$_{2}$H$^{+}$ derived 
using data from the literature at locations where line characteristics 
were reported are consistent with those derived by Bergin et al.

As shown by Bergin et al., fully-sampled maps from single-dish telescopes
can more effectively determine abundances within B68.  Further maps of B68 
can be now easily obtained using the new multi-beam focal-plane arrays on 
current single-dish telescopes.  Furthermore, the upcoming generation of 
millimeter interferometers will have the sensitivity to provide additional 
data for combined maps of even higher resolution.  With these data, more 
accurate molecular abundances will be determined for B68.  For example, it 
will be interesting to determine if abundances of CS, HCO$^{+}$, NH$_{3}$, 
H$_{2}$CO, and C$_{3}$H$_{2}$ also vary with extinction, like $^{13}$CO, 
C$^{18}$O, or N$_{2}$H$^{+}$.  Similar abundance determinations across 
other isolated dense cores may also be possible, provided their density 
structures can be similarly well-defined.

\acknowledgements{We thank Barry Turner, our referee, for insightful 
comments that improved this paper.  In addition, we thank Leo Blitz for 
allowing this project to be pursued with the BIMA millimeter array.  
JD's research in Berkeley was supported by the Radio Astronomy Laboratory.  
MRH's research in Berkeley was supported by the Miller Institute for 
Basic Research in Science.  We also thank Charles J. Lada, Jo\~ao Alves, 
Tracy Huard, and Jon Swift for their help.  This research has made use 
of the SIMBAD database and Aladin, both operated by CDS, Strasbourg, 
France.}

\clearpage

\figcaption{ESO MAMA $R$-band image of B68 (from CDS/Aladin).  
The position of maximum visual extinction determined by ALL01, i.e., the 
extinction peak, is denoted by a diamond.  The central pointing position 
of the BIMA observations is shown as a cross and the FWHM of the BIMA 
primary beams at 89.2 GHz (HCO$^{+}$ 1--0) and 110.2 GHz ($^{13}$CO 1--0) 
are shown respectively as the outer and inner dashed circles.  The 
positions and FWHM beam sizes of single-dish line observations from 
the literature are shown respectively as triangles and dotted circles.  
From north to south, these positions are from Launhardt et al.\/ (1998; 
CS 2--1), Benson et al. (1998; N$_{2}$H$^{+}$ 1--0 and C$_{3}$H$_{2}$ 
2$_{12}$--1$_{01}$), Wang et al.\/ (1995; C$^{18}$O 2--1 and H$_{2}$CO 
3$_{21}$--2$_{11}$), and Lemme et al.\/ (1996; NH$_{3}$ (1,1)).  The 
``X" denotes the position of B68 from Clemens \& Barvainis (1988).  The 
error ellipse of the nearby far-infrared point source IRAS 17194-2351 
is located south of B68 off the side of the image shown.  The scale bar 
size assumes a distance to B68 of 125 pc. \label{fig1}} 

\clearpage

\renewcommand{\arraystretch}{.6}

% Table 1
\begin{deluxetable}{ccccc}
\tablewidth{0pc}
\tablecaption{Summary of Line Observations of B68}
\tablehead{
& & \colhead{Beam FWHM\tablenotemark{a}} & \colhead{Extinction Peak Offset\tablenotemark{b}} \\
\colhead{Line}  & \colhead{Observatory} & \colhead{(\as\ ~$\times$~~\as)} & \colhead{(\as)} & \colhead{Reference\tablenotemark{~}}
} 
\startdata
$^{13}$CO 1--0                    &  BIMA    & 17.1 $\times$ 4.7 & 4.03 & 1 \\
C$^{18}$O 1--0                    &  BIMA    & 17.7 $\times$ 4.5 & 4.03 & 1 \\
C$^{18}$O 2--1                    &   CSO    &  30 $\times$ 30   & 59.3\tablenotemark{c} & 2 \\
CS 2--1                           &  FCRAO   &  46 $\times$ 46   & 66.2\tablenotemark{c} & 3 \\
HCO$^{+}$ 1--0                    &  BIMA    & 18.5 $\times$ 7.0 & 4.03 & 1 \\
HOC$^{+}$ 1--0                    &  BIMA    & 18.9 $\times$ 7.0 & 4.03 & 1 \\
N$_{2}$H$^{+}$ 1--0               &  BIMA    & 17.7 $\times$ 6.6 & 4.03 & 1 \\
N$_{2}$H$^{+}$ 1--0               & Haystack &  18 $\times$ 18   & 16.3 & 4 \\
NH$_{3}$ (1,1)                    &Effelsberg&  40 $\times$ 40   & 54.5 & 5 \\
H$_{2}$CO 3$_{12}$--2$_{11}$      &   CSO    &  30 $\times$ 30   & 59.3 & 2 \\
C$_{3}$H$_{2}$ 2$_{12}$--1$_{01}$ & Haystack &  18 $\times$ 18   & 16.3 & 4 \\
\enddata
\tablenotetext{~}{References --- (1) this work, (2) Wang et al.\/ 1995, 
(3) Launhardt et al.\/ 1998, (4) Benson, Caselli \& Myers 1998, (5) Lemme 
et al.\/ 1996.} 
\tablenotetext{a}{Synthesized beam FWHM for BIMA data, observed beam FWHM
for single-dish data reported by authors.}
\tablenotetext{b}{Offset of data from the extinction peak.}
\tablenotetext{c}{Position of peak emission from map, where line 
characteristics were reported by authors.}
\end{deluxetable}

\renewcommand{\arraystretch}{.6}

% Table 2 
\begin{deluxetable}{ccccc}
\tablewidth{0pc}
\tablecaption{Abundances of Various Molecular Clouds}
\tablehead{
& & \colhead{Translucent} \\
\colhead{Species} & \colhead{B68\tablenotemark{a}} & \colhead{Clouds\tablenotemark{b}} & \colhead{TMC-1\tablenotemark{c}} & \colhead{L134N\tablenotemark{c}}
}
\startdata
$^{13}$CO      & $<$1.3(-7)  & \nodata & 8.9(-7)\tablenotemark{d} & 8.9(-7)\tablenotemark{d} \\ 
C$^{18}$O      & $<$1.3(-7)  & \nodata & 1.6(-7)\tablenotemark{d} & 1.6(-7)\tablenotemark{d} \\ 
C$^{18}$O      &   3.0(-8)   & \nodata & 1.6(-7) & 1.6(-7) \\ 
CS 2--1        &   4.0(-10)  & 1.1(-9) &  1(-8)  &  1(-9)  \\ 
HCO$^{+}$      & $<$1.4(-10) &  2(-9)  &  8(-9)  &  8(-9)  \\ 
HOC$^{+}$      &   \nodata   & \nodata & \nodata & \nodata \\ 
N$_{2}$H$^{+}$ & $<$1.3(-10) &  1(-9)  &  5(-10) &  5(-10) \\ 
N$_{2}$H$^{+}$ &   2.0(-11)  &  1(-9)  &  5(-10) &  5(-10) \\ 
NH$_{3}$       &   7.0(-10)  & 2.1(-8) &  2(-8)  &  2(-7)  \\
H$_{2}$CO      &   4.0(-10)  & 6.3(-9) &  2(-8)  &  2(-8)  \\ 
C$_{3}$H$_{2}$ &   1.2(-11)  & 3.6(-8) &  3(-8)  &  2(-9)  \\ 
\enddata
\tablenotetext{a}{Abundances derived at positions listed in Table 1
assuming Bonnor-Ebert model of ALL01 and T = 16~K.  Upper limits are 
derived from interferometer non-detections while values are derived 
from single-dish detections described in the literature.  Using the 
temperature profile suggested by Zucconi, Walmsley, \& Galli (2001) 
increases abundances by factors of 2--3.} 
\tablenotetext{b}{Translucent cloud abundances from Turner (2000) are 
defined for a fiducial cloud of constant abundance with edge-to-center 
visual extinction of 2.0.  The values listed are averages from hydrostatic
equilibrium polytropic models and constant-density models.}
\tablenotetext{c}{Cold dense cloud abundances from Ohishi, Irvine, 
\& Kaifu (1992), assuming N(H$_{2}$) = 10$^{22}$ cm$^{-2}$.}
\tablenotetext{d}{Derived from $^{12}$CO abundance assuming 
$^{12}$C/$^{13}$C = 90 or $^{16}$O/$^{18}$O = 500.}
\tablecomments{$a(-b)$ denotes $a \times 10^{-b}$}
\end{deluxetable}

\clearpage

\begin{figure}
\vspace{7.25in}
\hspace{-0.25in}
\includegraphics{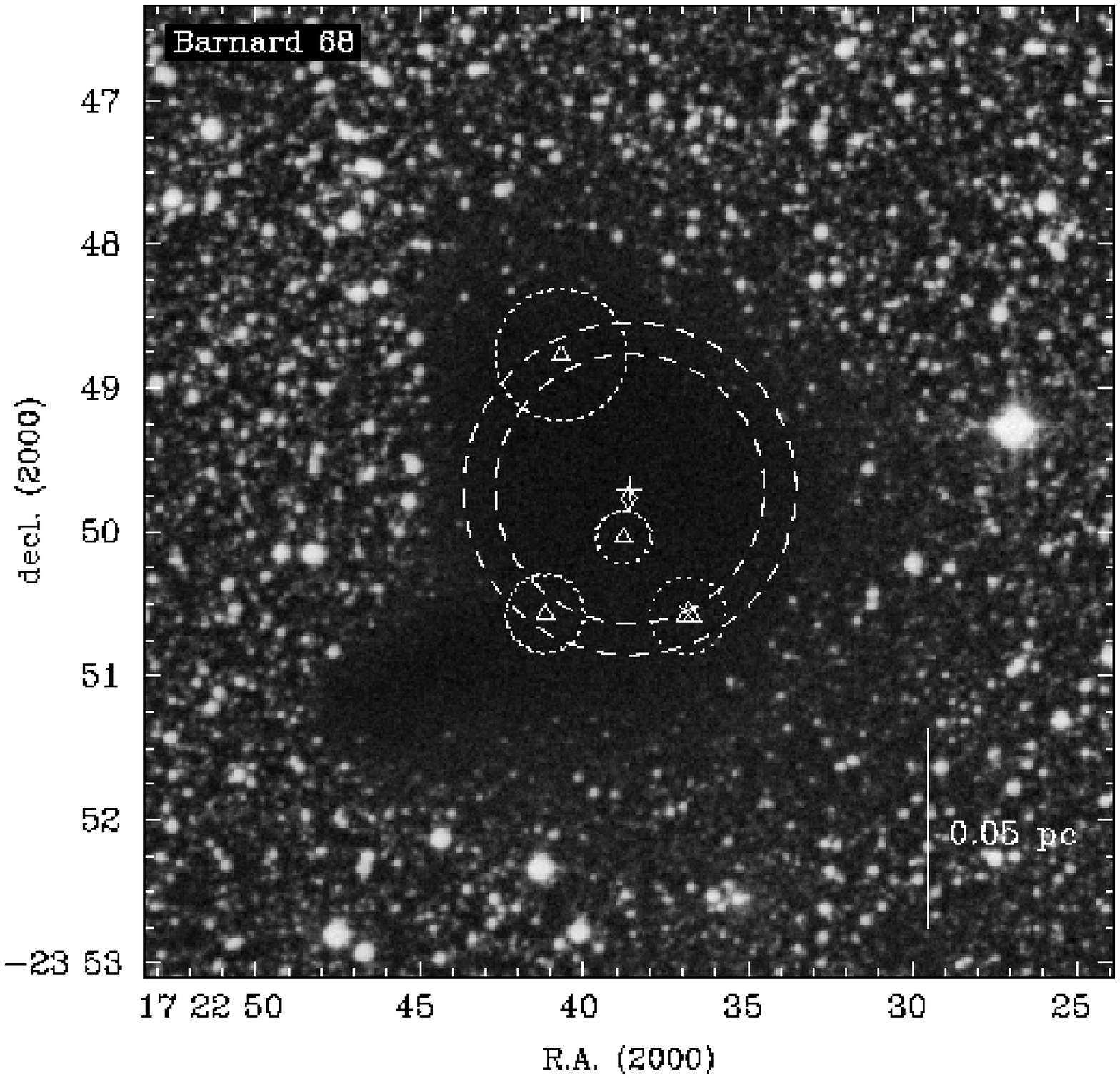}
\vspace{-2.25in}
\end{figure}

\end{document}